\documentstyle[12pt,a4]{article}

\begin{document}
\renewcommand {\thepage} { }
\renewcommand {\thefootnote} {\fnsymbol{footnote}}
\setcounter {page} {0}
\setcounter {footnote} {0}
\vspace*{1cm}
\noindent FSUJ TPI QO-03/96
\begin{flushright}
March, 1996
\end{flushright}
\vspace{5mm}
\begin{center}{\Large \bf
Homodyne detection for measuring \\[.5ex] internal quantum correlations
of optical pulses
}\end{center}
\vspace{5mm}
\begin{center}
T. Opatrn\'{y}\footnote{Permanent address:
Palack\'{y} University, Faculty of Natural Sciences, Svobody 26,
77146 Olomouc, Czech Republic}, D.-G. Welsch \\
Friedrich-Schiller Universit\"at Jena,
Theoretisch-Physikalisches Institut \\
Max-Wien Platz 1, D-07743 Jena, Germany \\[1ex]
W.~Vogel \\
Universit\"at Rostock, Fachbereich Physik \\
Universit\"atsplatz~3, D-18051 Rostock, Germany 
\end{center}
\vspace{3cm}

\begin{center}{\bf Abstract}\end{center}
A new method is described for determining the quantum
correlations at different times in optical pulses by using
balanced homodyne detection. The signal pulse and sequences
of ultrashort test pulses are superimposed, where for
chosen distances between the test pulses
their relative phases and intensities are varied
from measurement to measurement. The correlation statistics of
the signal pulse is obtained from the time-integrated
difference photocurrents measured.

\vspace{1cm}

\vfill
\newpage
\renewcommand {\thepage} {\arabic{page}}
\setcounter {page} {1}
Methods for measuring the quantum statistics
of optical
fields have been of increasing interest. One of the most fruitful
experimental methods has been balanced homodyne detection
\cite{Yuen1} 
including
homodyne correlation measurements 
\cite{Ou1}. 
In particular,
combining a single-mode signal field and a strong local oscillator
(LO) by a 50\%:50\% beam splitter and measuring the interfering
fields in the two output channels by means of two photodetectors,
the difference-count statistics yields directly the field-strength
statistics of the signal mode for a certain phase parameter. Since
knowledge of the probability distributions of the field strengths for all
phases within a $\pi$-interval is equivalent to knowledge of the
quantum state \cite{KVogel1}, the field-strength distributions
can be used to reconstruct the quantum state in terms of the
density matrix or other representative quantities, such as
phase-space functions.

In the pioneering experiments by Smithey,
Beck, Raymer, and Faridani \cite{Smithey1} the Wigner function
was determined from the measured field-strength distributions by
means of inverse Radon transform and the density-matrix
elements in a field-strength basis were then obtained as Fourier
transforms of the Wigner function (see also Ref.~\cite{Smithey2}).
The method also called optical homodyne tomography (OHT) has been
improved in order to obtain the density matrix in a more direct
way 
\cite{Kuehn1}. 
In particular, the problem of
direct sampling of the density matrix in the photon-number
basis has been studied in a number of papers
\cite{Ariano2}.
The method can be extended to the detection of multimode light by
increasing the number of the input (and output) channels of
the apparatus. The problem of reconstruction of the quantum state
of two correlated modes was first considered in Ref.~\cite{Raymer1},
and an analysis of the determination of the quantum state of a
correlated $N$-mode field by multiport homodyning is given in
Refs.~\cite{Kuehn2}.

In the experiments reported in Ref.~\cite{Smithey1} optical pulses
are used and the time-integrated difference-count statistics is
measured. Hence, a single-mode density matrix is reconstructed from
the measured data, which only yields information on the quantum statistics
of the signal pulse as a whole, but does not contain any information about
its internal quantum statistics. Recently, Munroe, Boggavarapu,
Anderson, and Raymer \cite{Munroe1} have performed OHT experiments
using LO pulses
that are short compared with the signal pulses.
In this way they have been able to obtain the photon-number statistics
of a signal pulse at different times in the pulse.

In this paper we propose a method for measuring also the internal
correlation statistics of optical pulses, which applies to both classical
and nonclassical light. In particular, the method
can be used to study correlation-assisted coherence properties
in an experiment analogous to Young's interference experiment, but
with a ``temporal double slit'' inside a pulse. Another example
is the measurement of the photon-number correlation at different
times in the pulse. It is worth noting that even in
classical optics the measurement of the internal amplitude-phase
structure of pulses is not trivial and has been a subject of current
interest 
\cite{Diels1}.

\begin{figure}[h]
\unitlength1cm
\begin{center}
\begin{picture}(8,10)
\put(1,3){\line(1,0){2.5}}
\put(1,7.5){\line(1,0){6.5}}
\put(5,9.5){\line(1,0){1}}
\put(5.5,10.5){\line(1,0){2.5}}
\put(8.25,7.5){\line(1,0){0.25}}
\put(8.25,10.5){\line(1,0){0.5}}

\put(5.5,4){\line(0,1){5.5}}
\put(7.5,7){\line(0,1){1}}
\put(5.5,10.25){\line(0,1){0.25}}
\put(8.5,7.5){\line(0,1){2.5}}

\put(7.5,7.5){\oval(1.5,1)[r]}
\put(5.5,9.5){\oval(1,1.5)[t]}

\put(8.5,10.5){\circle{1}}

\put(4.735,6.765){\line(1,1){1.5}}
\put(4.765,6.735){\line(1,1){1.5}}

\put(2,3.2){\oval(0.15,4)[t]}
\put(1.725,3.2){\oval(0.4,0.4)[br]}
\put(2.275,3.2){\oval(0.4,0.4)[bl]}

\put(5.7,6.6){\oval(2,0.15)[r]}
\put(5.7,6.875){\oval(0.4,0.4)[bl]}
\put(5.7,6.325){\oval(0.4,0.4)[tl]}

\put(5.7,6){\oval(0.8,0.15)[r]}
\put(5.7,6.275){\oval(0.4,0.4)[bl]}
\put(5.7,5.725){\oval(0.4,0.4)[tl]}

\put(5.7,5.5){\oval(1.6,0.15)[r]}
\put(5.7,5.775){\oval(0.4,0.4)[bl]}
\put(5.7,5.225){\oval(0.4,0.4)[tl]}

\put(3,7.5){\oval(3,0.4)[t]}

\put(7,5){\vector(0,1){0.75}}
\put(2.3,4){\vector(1,0){0.75}}
\put(3,8){\vector(1,0){0.75}}

\put(1.5,8){\makebox(1,1){{\large Signal}}}

\put(7.5,5){\makebox(1,1){{\large LO}}}

\put(3.5,2){\framebox(4,2){{\Large $\varphi _{k}$, $q_{k}$}}}

\put(6,4.5){\makebox{\shortstack{.\\.\\.\\.}}}

\end{picture}

\end{center}

\caption{
Scheme of the measurement. The signal pulse and a train of
strong local-oscillator (LO) pulses that are short
compared with the signal pulse are combined by a 50\%:50\% beam
splitter, and the time-integrated difference-count statistics
in the output channels is measured. The train of LO pulses
is produced interferometrically, so that the pulse distances,
relative phases, and intensities can be controlled.
\label{F1}
}
\end{figure}
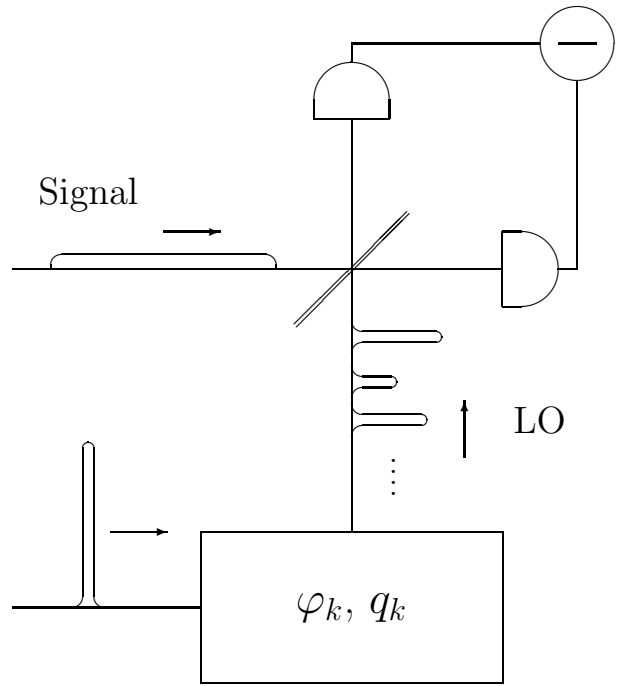

To measure the internal quantum statistics of an optical pulse,
an apparatus is desired that performs an appropriate mode decomposition
of the pulse without introduction of additional noise and measures
the quantum statistics of the correlated modes. 
This can be achieved by combining the signal
pulse, through a 50\%:50\% beam splitter, with a train of $N$ LO pulses
that are short compared with the signal pulse
and measuring the time-integrated difference-count statistics
in the two output channels of the beam splitter (Fig.\ref{F1}).
The signal pulse consists, in general, of a continuum of
monochromatic modes with photon destruction and creation
operators $\hat{b}(\omega)$ and $\hat{b}^\dagger(\omega')$,
respectively, $[\hat{b}(\omega),$ $\!\hat{b}^\dagger(\omega')]$
$\!=$ $\! \delta(\omega'\! -\! \omega)$.
Let us suppose that the 
strong LO pulses are prepared in coherent states,
so that the positive-frequency part of the $k$th pulse centered at
time $t_k$ can be described by a function $\gamma_{k} g_{k}(t)$,
where $g_{k}(t)$ is assumed to be normalized to unity and $\gamma_{k}$
is a complex number. We now consider the operators
\begin{equation}
\label{1}
\hat{a}_k = \int {d}\omega \, \tilde{g}_k^*(\omega) \, \hat{b}(\omega),
\end{equation}
where $\tilde{g}_k(\omega)$ is the Fourier transform of $g_k(t)$,
$\tilde g_k(\omega )$ $\!\equiv$ $(2\pi )^{-1/2}$ $\!\int {d}\omega\,
g_{k}(t) \exp(i\omega t ) $. It can be shown that when the LO pulses
$g_k(t)$ and $g_{k'}(t)$, $k$ $\!\neq$ $\!k'$, are (approximately)
nonoverlapping, then the operators $\hat{a}_k$ and $\hat{a}_{k'}^\dagger$
satisfy the standard bosonic commutation relation
$[\hat{a}_{k},\hat{a}_{k'}^{\dagger}]\!=\!\delta_{kk'}$.
Hence, the train of LO pulses can be used to introduce a set of
nonmonochromatic modes 
\cite{Titulaer1}
and ``probe'' the correlation statistics of
the signal pulse in terms of these modes. In particular,
measurement of the integrated difference-count statistics
can be shown to be equivalent to measurement of the 
sum of signal-pulse
field strengths
\begin{equation}
\hat{F} = \sum_k q_{k} \hat{F}_k(\varphi_k).
\label{2}
\end{equation}
Here,
\begin{equation}
\hat{F}_k(\varphi_k) = 2^{-1/2} \left(
\hat{a}_k e^{-i\varphi_k} + \hat{a}_k^\dagger e^{i\varphi_k}
\right)
\label{3}
\end{equation}
is the field strength associated with the $k$th pulse-like
nonmonochromatic mode centered at time $t_k$ in the signal pulse
and $q_{k}$ is a non-negative
real parameter describing the relative (square root of the) intensity
of the $k$th LO pulse. 
Measuring the distribution of the sum field strength (\ref{3})
for a sufficiently large set of values of $\varphi_k$ and $q_k$, we may
obtain the quantum statistics of the signal pulse within the frame of an 
$N$-mode density matrix.

The train of LO pulses may be obtained from one generating
LO pulse using interferometric methods, so that the distances
between the LO pulses, their intensities and relative phases
can be controlled. Let us further assume that the signal pulse
can be appropriately triggered by the generating LO. In this way,
the times $t_k$ in the signal pulse and the absolute values and
the relative phases of the $F_k$ can be controlled.
When the ``positions'' $t_k$ of the LO pulses are shifted towards
$t_k$ $\!+$ $\!\Delta t$, then the phases $\varphi_k$ are shifted
towards $\varphi_k$ $\!+$ $\!\omega_0\Delta t$, $\omega_0$ being
the carrier frequency of the signal and LO pulses. Since times of
the order of magnitude of $\omega_0^{-1}$ cannot be controlled, an
absolute phase control seems to be impossible when the signal and
LO pulses come from different sources. In this case, only
quantities averaged with respect to the unknown phase can be obtained.

Let us consider the simplest case when the signal pulse and two
short nonoverlapping LO pulses centered at the times $t_1$ and
$t_2$ in the signal pulse are superimposed. In this case, measurement
of the time-integrated difference-count statistics yields the statistics
of a sum of
two-mode field strengths of the signal pulse,
\begin{equation}
\label{4}
\hat{F} = \hat{F}(\varphi,\Delta\varphi,q) = \hat{F}_{1}(\varphi)
+ q\, \hat{F}_{2}(\varphi + \Delta \varphi).
\end{equation}
The measured moments of $\hat{F}$ are related
to the moments and correlation functions of the signal pulse at the two
times $t_1$ and $t_2$ in the pulse as
\begin{equation}
\big\langle{\hat{F}}^n\big\rangle
= \sum_{k=0}^n {n \choose k} q^{k}
\big\langle {\hat{F}_1}^{n-k}(\varphi)
{\hat{F}_2}^{k}(\varphi+\Delta\varphi)\big\rangle.
\label{5}
\end{equation}
Measuring $\langle{\hat{F}}^n\rangle$
for $n\!+\!1$ values of $q$, from Eq.~(\ref{5}) one obtains a set
of $n\!+\!1$ linear algebraic equations whose solution yields
the signal-pulse moments and correlation functions
$\langle {\hat{F}_1}^{n-k}(\varphi)
{\hat{F}_2}^{k}(\varphi+\Delta\varphi)\rangle$
for the chosen values of $\varphi$ and $\Delta \varphi$.
Varying $\varphi$ and $\Delta \varphi$, the procedure can be repeated
many times to obtain the $\varphi$ and $\Delta \varphi$ dependences
of the moments and correlation functions. 
In the limit when the number of measurements   
goes to infinity all the moments and correlations can be obtained, 
the knowledge of which is equivalent to knowledge of the two-mode density matrix.
Clearly, when the phase $\varphi$
cannot be controlled the phase-averaged (even) moments
\begin{equation}
\overline{\big\langle{\hat{F}}^{n}\big\rangle}
\equiv \frac{1}{2\pi}\int {d}\varphi \,
\big\langle{\hat{F}}^{n}\big\rangle
\label{5a}
\end{equation}
can only be measured and Eq.(\ref{5}) modifies to
\begin{eqnarray}
\label{5b}
\overline{\big\langle{\hat{F}}^n\big\rangle}
= \sum_{k=0}^n {n \choose k} q^{k}
\overline{\big\langle {\hat{F}_1}^{n-k}(\varphi)
{\hat{F}_2}^{k}(\varphi+\Delta\varphi)\big\rangle},
\end{eqnarray}
$n$ being even.

Let us briefly discuss the situation of nonperfect detection.
In this case, 
$\hat{a}_{k}$ must be replaced by $\eta\hat{a}_{k}$ $\!+$
$\sqrt{\eta(1\!-\!\eta)}\hat{c}_k$ (cf. Ref.~\cite{Yurke1}),
where $\eta$ is the detection efficiency ($\eta$ $\!\leq$ $\!1$)
and $\hat{c}_k$ and $\hat{c}_k^\dagger$ are bosonic noise
operators. 
The corresponding noise modes can be considered to be in vacuum states, so 
that for each of these modes the $m$th moment of the corresponding field strength 
is given by $
2^{-m/2}(m-1)!!$ for even $m$ and zero for odd $m$
[$(m\! -\! 1)!!$ $\! =$ $ \! 1\cdot 3 \cdots (m-3)(m-1)$ for $m\ge 2$;
$(-1)!!$ $\! \equiv$ $\! 1$].
We then find that in the sum in Eq.~(\ref{5b}) 
the replacement 
\begin{eqnarray}
\lefteqn{
\overline{\big\langle {\hat{F}_1}^{n-k}(\varphi)
{\hat{F}_2}^{k}(\varphi+\Delta\varphi)\big\rangle} \to
\sum_{l,m} {n\!-\!k \choose l} {k \choose m}
\eta ^{\frac{n+l+m}{2}} \!
\left( \frac{1\!-\!\eta}{2} \right)^{\!\frac{n-l-m}{2}}
}
\nonumber \\ && \hspace{15ex} 
\times \,
(n\!-\!k\!-\!l\!-\!1)!!\,(k\!-\!m\!-\!1)!! \,
\overline{\big\langle {\hat{F}_1}^{l}(\varphi)
{\hat{F}_2}^{m}(\varphi+\Delta\varphi)\big\rangle}
\label{5c}
\end{eqnarray}
\noindent
must be made.
When $k$ is even (odd) the $l$- and $m$-sums in Eq.~(\ref{5c}) run over the even (odd) 
integers between zero and $n$ $\!-$ $\!k$  
and between zero and $k$, 
respectively.
The replacement means that the reconstructed moments and correlations   
in Eq.~(\ref{5b})
are attenuated and ``contaminated'' by the moments of lower orders. 
Nevertheless, we can recurrently correct  
them
by subtracting 
the contaminating terms, because only lower-order terms are included.

To illustrate the method, let us briefly consider the
phase-averaged second- and forth-order moments of the measured
time-integrated difference-count statistics.
Information about the lowest-order correlation in the signal
pulse at two times $t_1$ and $t_2$ is obtained
from the second-order moment,
\begin{equation}
\overline{\big\langle{\hat{F}}^2\big\rangle}
= \overline{\big\langle{\hat{F}_1}^2(\varphi)\big\rangle}
+ q^{2}\,
\overline{\big\langle{\hat{F}_2}^2(\varphi)\big\rangle}
+ 2 q \overline{\big\langle {\hat{F}_1}(\varphi)
{\hat{F}_2}(\varphi+\Delta\varphi)\big\rangle}.
\label{6}
\end{equation}
Whereas $\overline{\big\langle{\hat{F}_1}^2(\varphi)\big\rangle}$
and $\overline{\big\langle{\hat{F}_2}^2(\varphi)\big\rangle}$
are closely related to the mean numbers of photons at the
two times in the signal pulse 
\begin{equation}
\label{6a}
\overline{\big\langle{\hat{F}_k}^2(\varphi)\big\rangle}
= 
\langle\hat{n}_k\rangle + \textstyle\frac{1}{2}
\end{equation}
($k$ $\!=$ $1,2$), the cross term
\begin{equation}
\overline{\big\langle {\hat{F}_1}(\varphi)
{\hat{F}_2}(\varphi+\Delta\varphi)\big\rangle}
= 
\textstyle\frac{1}{2} 
\big\langle\hat{a}_1^\dagger\hat{a}_2\big\rangle
e^{-i\Delta\varphi}
+ {\rm c.c.}
\label{7}
\end{equation}
as a function of the difference phase $\Delta \varphi$ can
be regarded as the second-order coherence function that
probes the effect of signal-pulse interference at a
``temporal double slit'' in the pulse \cite{Per}.
Choosing $q$ $\!=$ $\!1$ and measuring the second-order moment
$\overline{\langle{\hat{F}}^2\rangle}$ for a set of phase
differences $\Delta \varphi$ and $\Delta\varphi$ $\!+$ $\!\pi$, the
interference term is simply given by the difference of the measured
moments at the two phase differences.
In the case of nonperfect detection the 
photon-number  
terms are attenuated
and contaminated by a constant,
\begin{equation}
\overline{\big\langle{\hat{F}_k}^2(\varphi)\big\rangle}
\to
\eta ^{2}
\overline{\big\langle{\hat{F}_k}^2(\varphi)\big\rangle}
+ \textstyle\frac{1}{2}\eta (1-\eta),
\label{7a}
\end{equation} 
whereas the interference term is only attenuated,
\begin{equation}
\overline{\big\langle {\hat{F}_1}(\varphi)
{\hat{F}_2}(\varphi+\Delta\varphi)\big\rangle}
\to
\eta ^{2} 
\overline{\big\langle {\hat{F}_1}(\varphi)
{\hat{F}_2}(\varphi+\Delta\varphi)\big\rangle}.
\label{7b}
\end{equation}

Extending the measurement to the fourth-order moments
$\overline{\langle{\hat{F}}^4\rangle}$ for different values of $q$
and $\Delta \varphi$, the phase-averaged fourth-order moments and
and correlations of $\hat{F}_1(\varphi)$ and $\hat{F}_2(\varphi)$
can be obtained. The moments
$\overline{\langle{\hat{F}_1}^4(\varphi)\rangle}$
and $\overline{\langle{\hat{F}_2}^4(\varphi)\rangle}$ carry information
about the photon-number statistics (first- and second-order moments of
the photon numbers) of the signal pulse at two different times in
the pulse. 
The two-time correlations are given by
\begin{equation}
\overline{\big\langle{\hat{F}_1}^3(\varphi)
\hat{F}_2 (\varphi + \Delta \varphi) \big\rangle}
= 
\textstyle\frac{3}{2} \,
\overline{\big\langle\hat{F}_1(\varphi)
\hat{F}_2 (\varphi + \Delta \varphi) \big\rangle}
+  \textstyle\frac{3}{4}
\big(
\big\langle \hat{a}_{1}^{\dagger 2} \hat{a}_{1} \hat{a}_{2} \big\rangle
e^{-i \Delta \varphi}
+ \rm{c.c.}
\big),
\label{8}
\end{equation}
\begin{equation}
\overline{\big\langle\hat{F}_1(\varphi)
{\hat{F}_2}^3(\varphi + \Delta \varphi) \big\rangle}
= 
\textstyle\frac{3}{2} \,
\overline{\big\langle\hat{F}_1(\varphi)
\hat{F}_2(\varphi + \Delta \varphi) \big\rangle}
+  \textstyle\frac{3}{4}
\big(
\big\langle \hat{a}_{2}^{\dagger 2} \hat{a}_{2} \hat{a}_{1} \big\rangle
e^{i \Delta \varphi}
+ \rm{c.c.}
\big)
\label{9}
\end{equation}
with $\overline{\langle\hat{F}_1(\varphi)
\hat{F}_2(\varphi + \Delta \varphi) \rangle}$ from Eq.~(\ref{7}),
and
\begin{equation}
\label{10}
\overline{\big\langle {\hat{F}_{1}}^{2}(\varphi)
{\hat{F}_2}^{2} (\varphi + \Delta \varphi)
\big\rangle}
=  
\textstyle\frac{1}{4}
\left(
\big\langle \hat{a}_{1}^{\dagger 2} \hat{a}_{2}^{2} \big\rangle
e^{-i2 \Delta \varphi} + \rm{c.c.} \right)
+  
\big\langle \big( \hat n_{1} + \textstyle\frac{1}{2} \big)
\big( \hat n_{2} + \textstyle\frac{1}{2} \big) \big\rangle.
\end{equation} 
In particular, from Eq.~(\ref{10}) together with the mean 
photon numbers obtained from the second-order moments, 
Eq.~(\ref{6a}), the photon-number correlation at the 
times $t_1$ and $t_2$ in the signal pulse can be
determined, which for stationary ergodic fields is 
usually measured in the Hanbury-Brown and Twiss experiment.
When the detection efficiency is less than unity
the terms $\overline{\langle{\hat{F}_k}^4(\varphi)\rangle}$ 
changes as
\begin{equation}
\overline{\big\langle{\hat{F}_k}^4(\varphi)\big\rangle}
\to
\eta ^{4}
\overline{\big\langle{\hat{F}_k}^4(\varphi)\big\rangle}
+ 3 \eta ^{3} (1-\eta)
\overline{\big\langle{\hat{F}_k}^2(\varphi)\big\rangle}
+ \textstyle\frac{3}{4} \eta ^{2} (1-\eta) ^{2},
\label{10a}
\end{equation}
and with respect to the symmetrical cross term (\ref{10}) we find that
\begin{eqnarray}
\lefteqn{
\overline{\big\langle {\hat{F}_{1}}^{2}(\varphi)
{\hat{F}_2}^{2} (\varphi + \Delta \varphi)
\big\rangle}
\to
\eta ^{4}
\overline{\big\langle {\hat{F}_{1}}^{2}(\varphi)
{\hat{F}_2}^{2} (\varphi + \Delta \varphi)
\big\rangle}
}
\nonumber \\ && \hspace{5ex}
+ \textstyle\frac{1}{2} \eta ^{3} (1-\eta) \big[ 
\overline{\big\langle{\hat{F}_1}^2(\varphi)\big\rangle}
+ \overline{\big\langle{\hat{F}_2}^2(\varphi)\big\rangle}
\big]
+ \textstyle\frac{1}{4} \eta ^{2} (1-\eta)^{2} .
\end{eqnarray}

The inclusion in the analysis of higher-order moments is straightforward.
In summary, the proposed method may serve as a useful tool for determining 
the correlation statistics of optical pulses, with special emphasis on 
nonclassical light. The application to pulses becomes feasible since only 
ensemble averaging is used. The other advantage of the method is that it 
also applies to weak quantum light, because  homodyning (with strong LO) 
``amplifies'' the signal. In particular, the method can be used to
obtain the correlation statistics of pulses of the type
used in Ref.~\cite{Munroe1} for studying the photon-number 
statistics at different times in a pulse.  
In the paper we have restricted attention
to the determination of moments and correlations at different times in the
signal pulse. The problem of reconstruction of the associated density matrices
will be considered in a forthcoming paper.

\vspace{1ex}

This work was supported by the Deutsche Forschungsgemeinschaft. One of us (T.O.)
is grateful to J.~Pe\v{r}ina for discussion and acknowledges a support of an
internal grant of the Palack\'{y} University.

\vspace{5ex}

\noindent
%
%
%
%



\newpage

\begin{thebibliography}{99}
\bibitem{Yuen1}
H. P. Yuen and J.H. Shapiro,
IEEE Trans. Inform. Theor. {\bf 26}, 78 (1980);
H. P. Yuen and V.W.S. Chan,
Opt. Lett. {\bf 8}, 177 (1983);
N. G. Walker,
J. Mod. Opt. {\bf 34}, 15 (1987);
S. Braunstein,
Phys. Rev. A{\bf 42}, 474 (1990);
W. Vogel and J. Grabow,
Phys. Rev. A{\bf 47}, 4227 (1993).

\bibitem{Ou1}
Z.Y. Ou, C.K. Hong, and L. Mandel,
Phys. Rev. A{\bf 36}, 192 (1987);
W. Vogel,
Phys. Rev. Lett. {\bf 67}, 2450 (1991);
Phys. Rev. A{\bf 51}, 4160 (1995).

\bibitem{KVogel1}
K. Vogel and H. Risken,
Phys. Rev. A{\bf 40}, 2847 (1989).

\bibitem{Smithey1}
D.T. Smithey, M. Beck, M.G. Raymer, and A. Faridani,
Phys. Rev. Lett. {\bf 70}, 1244 (1993).

\bibitem{Smithey2}
D.T. Smithey, M. Beck, J. Cooper, M. G. Raymer, and A. Faridani,
Phys. Scr. T{\bf 48}, 35 (1993).

\bibitem{Kuehn1}
H. K\"{u}hn, D.-G. Welsch, and W. Vogel,
J. Mod. Opt. {\bf 41}, 1607 (1994);
W. Vogel and D.-G. Welsch,
{\em Lectures on Quantum Optics} (Akademie-Verlag, Berlin, 1994);
G.M. D'Ariano, C. Machiavello, and M.B.A. Paris,
Phys. Rev. A{\bf 50}, 4298 (1994);
Phys. Lett. {\bf 195}A, 31 (1994).

\bibitem{Ariano2}
G.M. D'Ariano,
Quantum Semiclass. Opt. {\bf 7}, 693 (1995);
G.M. D'Ariano, U. Leonhardt, and H. Paul,
Phys. Rev. A{\bf 52}, R1881 (1995);
U. Leonhardt, H. Paul, and G.M. D'Ariano,
Phys. Rev. A{\bf 52}, 4899 (1995);
H. Paul, U. Leonhardt, and G.M. D'Ariano,
Acta Phys. Slov. {\bf 45}, 261 (1995);
Th. Richter,
Phys. Lett. A, in press;
U. Leonhardt, M. Munroe, T. Kiss, M.G. Raymer, and Th. Richter,
Opt. Commun., in press.

\bibitem{Raymer1}
M.G. Raymer, D.T. Smithey, M.Beck, M. Anderson, and D.F. McAlister,
Third International Wigner Symposium (Oxford, 1993).

\bibitem{Kuehn2}
H. K\"{u}hn,  D.-G. Welsch, and W. Vogel,
Phys. Rev. A{\bf 51}, 4240 (1995);
W. Vogel and D.-G. Welsch,
Acta Phys. Slov. {\bf 45}, 313 (1995);
A. Zucchetti,  W. Vogel, and D.-G. Welsch,
Phys. Rev. A, in press.

\bibitem{Munroe1}
M. Munroe, D. Boggvarapu, M.E. Anderson, and M.G. Raymer,
Phys. Rev. A{\bf 52}, R924 (1995).

\bibitem{Diels1}
J.-C. Diels, J.J. Fontaine, I.C. McMichael, and F. Simoni,
Appl. Opt. {\bf 24}, 1270 (1985);
K. Naganuma, K. Mogi, and H. Yamada,
IEEE J. Quantum Electron. {\bf 25}, 1225 (1989);
Appl. Phys. Lett. {\bf 54}, 1201 (1989);
J.L.A. Chilla and O.E. Martinez,
IEEE J. Quantum Electron. {\bf 27}, 1228 (1991);
C. Yan and J.C. Diels,
J. Opt. Soc. Am. B{\bf 8}, 1259 (1991);
D.J. Kane and R. Trebino,
Opt. Lett. {\bf 18}, 823 (1993);
M. Beck, M.G. Raymer, I.A. Walmsley, and V. Wong,
Opt. Lett. {\bf 18}, 2041 (1993);
K.W. DeLong, R. Trebino, and D.J. Kane,
J. Opt. Soc. Am. B{11}, 1595 (1994).

\bibitem{Titulaer1}
U.M. Titulaer and R.J. Glauber,
Phys. Rev. {\bf 145}, 1041 (1966).

\bibitem{Yurke1}
B. Yurke and D. Stoler,
Phys. Rev. A {\bf 36}, 1955 (1987).



\bibitem{Per}
For ordinary second-order coherence see, e.g.,
J. Pe\v{r}ina, {\em Coherence of Light\/}
(2nd. ed., Reidel, Dordrecht, 1985).



\end{thebibliography}
\end{document}